%% file: main.tex
\documentclass[manuscript,screen,sigconf]{acmart}
\usepackage{multirow}
\usepackage{natbib}
\usepackage{graphicx}
\usepackage{todonotes}
\usepackage{balance}
\setuptodonotes{color=blue!30}
\AtBeginDocument{%
  \providecommand\BibTeX{{%
    \normalfont B\kern-0.5em{\scshape i\kern-0.25em b}\kern-0.8em\TeX}}}

\copyrightyear{2024}
\acmYear{2024}
\setcopyright{acmlicensed}\acmConference[ICER '24 Vol. 1]{ACM Conference on International Computing Education Research V.1}{August 13--15, 2024}{Melbourne, VIC, Australia}
\acmBooktitle{ACM Conference on International Computing Education Research V.1 (ICER '24 Vol. 1), August 13--15, 2024, Melbourne, VIC, Australia}
\acmDOI{10.1145/3632620.3671122}
\acmISBN{979-8-4007-0475-8/24/08}

\acmConference[ICER'24]{20th ACM Conference on International Computing Education Research (ICER 2024)}{Tue 13 - Thu 15 August 2024}{Melbourne, Victoria, Australia}

\usepackage{hyperref}
\begin{document}

\title{Teaching Digital Accessibility in Computing Education: Views of Educators in India}

\input{macros}
\author{P D Parthasarathy}\email{p20210042@goa.bits-pilani.ac.in}\orcid{0000-0002-8723-2407}
\affiliation{%
\institution{Department of Computer Science and Information Systems, BITS Pilani, KK Birla Goa Campus,}
\city{}\state{Goa}
\country{India}
\postcode{403726}
 }

\author{Swaroop Joshi}\email{swaroopj@goa.bits-pilani.ac.in}\orcid{0000-0003-4536-2446}
\affiliation{%
\institution{Department of Computer Science and Information Systems, BITS Pilani, KK Birla Goa Campus,}
\city{}\state{Goa}
\country{India}
\postcode{403726}
}



\renewcommand{\shortauthors}{}

\begin{abstract}
    \input{abstract}
\end{abstract}




\maketitle

\input{intro}

\input{relatedWork}

\input{method}

\input{findings}

\input{discussion}

\section*{Acknowledgments}

We thank BITS Pilani for supporting this work under grant number GOA/ACG/2021-2022/Nov/05. We also thank anonymous reviewers for their valuable suggestions on our initial draft.






\end{document}

%% file: macros.tex

\newcommand{\hiddenauthor}[1]{\author{Anonymous Author#1}\email{anon#1@university.edu}
\affiliation{%
  \institution{Anon Institution}
  \city{Anon City}\state{Anon State}
  \country{Country}
  \postcode{12345}
}}

\newcommand{\pwd}{person with disabilities}
\newcommand{\pwds}{persons with disabilities}
\newcommand{\pc}{\%}

%% file: abstract.tex

In recent years, there has been rising interest from both governments and private industry in developing software that is accessible to all, including people with disabilities.
However, the computer science (CS) courses that ought to prepare future professionals to develop such accessible software hardly cover topics related to accessibility.
While there is growing literature on incorporating accessibility topics in computing education in the West, there is little work on this in the Global South, particularly in India, which has a large number of computing students and software professionals.
In this \emph{replication report}, we present (A) our findings from a replication of surveys used in the US and Switzerland on who teaches accessibility and barriers to teaching accessibility and (B) a qualitative analysis of perceptions of CS faculty in India about digital accessibility and teaching accessibility.
Our study corroborates the findings of the earlier surveys: very few CS faculty teach accessibility, and the top barriers they perceive are the same. The qualitative analysis further reveals that the faculty in India need training on accessibility concepts and disabilities sensitization, and exposure to existing and ongoing CS education research and pedagogies. In light of these findings, we present recommendations aimed at addressing these challenges and enhancing the integration of accessibility into computing education.

%% file: intro.tex

\section{Introduction}
\label{sec:intro}

According to the World Health Organization's 2011 report, there are over 1 billion \pwds\ (PWDs)\footnote{While some prefer the term `disabled person' over `\pwd,' the latter is acceptable usage in India, for instance, in the 2016 Rights of Persons with Disabilities act, and elsewhere. We use `\pwd' or the abbreviation `PWD' throughout this paper.} 
across the world~\citep{worldhealthorganizationWorldReportDisability2011}. 
Digital accessibility refers to designing and developing digital products, such as websites, software, and applications, to ensure they can be easily accessed and used by everyone, including individuals with disabilities. This involves ensuring that digital content is perceivable, operable, understandable, and robust for all users, regardless of their physical or cognitive abilities. Throughout this paper we use the term accessibility or a11y to refer to digital accessibility.

Although there has been a growing awareness of and demand for accessibility in the software industry in recent years, only 2\% of industry leaders report that they find it easy or very easy to recruit candidates with required accessibility skills~\citep{BridgingAccessibleTechnology2023}.
On the other hand, software professionals feel they did not learn accessibility-related technologies and skills in college-level computer science (CS) courses~\citep{patelWhySoftwareNot2020}.
Indeed, very few CS faculty teach these topics~\citep{shinoharaWhoTeachesAccessibility2018,soaresguedesHowAreWe2020}, although accessibility and Inclusive Design are CS Core topics in ACM/IEEE/AAAI Computer Science Curricula 2023 ~\citep{10.1145/3664191}.
As a result, most real-world software lacks accessibility. For instance,
\begin{itemize}
  \item More than 96\% of the world's top 1 million home pages have some accessibility defects~\citep{WebAIMWebAIMMillion2023}, and
  \item 97\% of nearly 500 popular Android apps across 23 business categories have accessibility violations~\citep{yanCurrentStatusAccessibility2019}.
\end{itemize}


The Global South, particularly India, plays an important role in the world of software development: 
On StackOverflow's 2022 developer survey~\citep{StackOverflow20222022}, India is second only to the US when ranked by the population of developers, which, according to some studies~\citep{liebyWorldwideProfessionalDeveloper2019}, it is likely to surpass by 2024.  A survey of software professionals in India highlights the urgent demand for accessibility education in the industry \cite{parthasarathyExploringNeedAccessibility2024}. 
Higher education institutes in India have over 2 million undergraduates enrolled in CS and related courses~\citep{departmentofhighereducationgovernmentofindiaAllIndiaSurvey2021}, who will continue to feed this growth.
Moreover, there are interesting dynamics concerning accessibility, disabilities, and inclusion: the enrollment of students with disabilities in higher education is extremely low (0.19\%)~\citep{departmentofhighereducationgovernmentofindiaAllIndiaSurvey2021} and, unlike the Global North, very few institutions have disabilities services offices on campuses.
Due to this, 
awareness about 
accessibility in general (not just digital accessibility) among CS teachers and students in this region
is likely to be low.
%
%
Any meaningful effort towards improving digital accessibility will be incomplete without improving the status of accessibility education in the Global South. As a first step towards this, we studied the perceptions of college-level CS faculty in India on digital accessibility since they are most likely to teach these topics to future software developers.

In 2018, \citeauthor{shinoharaWhoTeachesAccessibility2018}~\citep{shinoharaWhoTeachesAccessibility2018} conducted a survey of CS faculty in the US to understand the challenges CS educators face in teaching accessibility.
The top two barriers were: (a) the CS faculty felt accessibility is not a core part of the CS curriculum, and (b) they did not know enough to teach it.
Similar results were found in Switzerland after replication of the survey in 2020~\citep{soaresguedesHowAreWe2020}.
We replicated these surveys to investigate the following research questions:

\begin{description}
  \item[RQ1] Who is teaching accessibility?
  \item[RQ2] What barriers do faculty see to teaching accessibility?
\end{description}

In India, disability etiquette remains relatively low \cite{edwardrajPerceptionsIntellectualDisability2010, WorldBankDisabilityReport2009}, and educational instructors themselves often lack awareness of disability and accessibility \cite{Das2013InclusiveEI, TanejaJohansson2021EducationOC, 10.1371/journal.pone.0290016}.
Despite laws such as the Rights of Persons with Disabilities (RPD) Act of 2016 by the Government of India \cite{ministryofsocialjusticeandempowermentgovernmentofindiaRightsPersonsDisabilities2016} and initiatives like Sugamya Bharat, which aim to promote accessibility and inclusion, implementation and awareness at the educational level remain inadequate. Moreover, the IS 17802, a standard issued by the Bureau of Indian Standards outlining digital accessibility guidelines, suffers from inadequate awareness, adoption, and enforcement. Due to this unique context of India, we decided to explore further the views of the faculty on accessibility and the resources they need to integrate accessibility topics into their teaching. We added the following qualitative research questions:

\begin{description}
  \item[RQ3] What are the faculty members' perceptions of digital accessibility?
  \item[RQ4] What support do CS faculty seek to integrate accessibility into their teaching?
\end{description}

RQ1 and RQ2 mirror those examined in previous studies conducted in the US~\citep{shinoharaWhoTeachesAccessibility2018} and Switzerland~\citep{soaresguedesHowAreWe2020}, employing a similar survey instrument to address them. Additionally, we introduced RQ3 and RQ4, where we qualitatively analyzed responses from open-ended 1-1 interviews to explore these inquiries further, marking a novel contribution to the existing literature.

In the rest of the paper, we review some related work (Sec.~\ref{sec:relatedwork}), describe the survey and interview protocol and data analysis methods (Sec.~\ref{sec:method}), present our findings (Sec.~\ref{sec:findings}), discuss our results (Sec.~\ref{sec:discussion}), and conclude with our study's limitations and the way ahead (Sec.~\ref{sec:limitations}).


%% file: relatedWork.tex
\section{Related Work}
\label{sec:relatedwork}



We first summarize the 2018 survey of CS faculty in the US~\citep{shinoharaWhoTeachesAccessibility2018}:
Out of the 1,857 respondants from 318 US institutions, about a fifth reported teaching accessibility.
Those who teach accessibility are likely to be female, to have human-computer interaction (HCI) expertise, and to know someone with a disability.
The main barriers to teaching accessibility are the perceived incompatibility with core CS learning outcomes and a lack of faculty preparedness in accessibility.

The replication study from Switzerland~\citep{soaresguedesHowAreWe2020} reports on 56 responses from 21 institutes. Nine of these were college administrators. Only 8 out of the 49 (16.3\%) CS faculty reported teaching accessibility.
They found the same barriers as the US study. 
They also investigated how the administrative staff were dealing with accessibility.
Since it does not directly relate to teaching accessibility, we did not include this research question in our study.

Existing work on teaching accessibility includes a variety of efforts ranging from offering dedicated courses on accessibility and assistive technology~\citep{kurniawanGeneralEducationCourse2010, matauschAssistecUniversityCourse2006, waldDesign10Credit2008} to 
incorporating accessibility themes into pre-existing courses such as web development~\citep{freireAccessibilityWebMultimedia2013, rosmaitaAccessibilityFirstNew2006, wangHolisticPragmaticApproach2012},
software engineering~\citep{el-glalyTeachingAccessibilitySoftware2020, ludiIntroducingAccessibilityRequirements2007, rossAccessibilityFirstclassConcern2017},
programming fundamentals~\citep{cohenAccessibilityIntroductoryComputer2005, jiaInfusingAccessibilityProgramming2021}, 
artificial intelligence~\citep{tsengExplorationIntegratingAccessibility2022}, 
and mobile app development~\citep{bhatiaIntegratingAccessibilityMobile2023}.
In a research that is not confined to a particular course, \citeauthor{lewthwaiteResearchingPedagogyDigital2023}~\citep{lewthwaiteResearchingPedagogyDigital2023} report accessibility educators may not make use of existing and ongoing research based on learning theory and pedagogy.

\citeauthor{bakerSystematicAnalysisAccessibility2020}~\citep{bakerSystematicAnalysisAccessibility2020} did a systematic literature review of 51 papers on integrating accessibility in CS courses. They found three main integration approaches: (a) specialized courses on accessibility, (b) accessibility as a theme in otherwise traditional CS courses like web development, and (c) accessibility as a single module in an otherwise traditional CS course.
They observed four broad categories of accessibility learning outcomes: 
 \vspace{-4pt}
 \begin{itemize}
   \item Empathy towards potential users who need accessibility features in the software they use,
   \item Awareness --- laws, ethics, assistive technology, etc.,
   \item Technical knowledge --- WCAG guidelines and testing tools, and
   \item Opportunities in the field of accessibility in the software industry.
 \end{itemize}\vspace{-4pt}

\citeauthor{coverdaleTeachingAccessibilityShared2022}~\citep{coverdaleTeachingAccessibilityShared2022}  conducted a qualitative investigation involving 30 educators from academia and the workplace. Their study advocates for interdisciplinary education and training across roles and disciplines, emphasizing the mutual learning between academia and the workplace. \citeauthor{elglalyHCINeedAccessibility2024a}~\citep{elglalyHCINeedAccessibility2024a} contend that confining accessibility within Human-Computer Interaction (HCI) overlooks its broader significance across the Computer Science (CS) curriculum and lacks the necessary framework to support CS educators adequately. They propose a series of knowledge units comprising topics and exemplar learning objectives for an Accessibility Knowledge Area in Computing Education. 

Recently, to help educators integrate accessibility into core CS courses, a comprehensive textbook has been published by \citeauthor{TeachingAccessibleComputing2024}~\citep{TeachingAccessibleComputing2024}. This textbook is a practical resource for instructors aiming to incorporate accessibility principles and practices into their teaching curriculum.

%% file: method.tex

\section{Method}
\label{sec:method}

\subsection{The survey}
\label{sec:method:survey}

\subsubsection{Participants}

Like the US and Switzerland studies, we surveyed faculty in CS and related fields (Computer Applications, Computer Engineering, Information Technology, Information Systems, and other interdisciplinary computing departments) at 4- or 3-year colleges and universities across India.
We manually identified institutes offering relevant programs from University Grants Commission's 2021 annual report~\cite{UGCAnnualReport2022}
and curated a list of email addresses of CS faculty from the official websites of these institutes.
We used an internally developed web-scraping script and manually scrutinized its results. This finally yielded a total of 3,689 faculty email addresses from 488 institutes. 

\subsubsection{Survey Instrument}

We adapted the survey instrument developed by \citeauthor{shinoharaWhoTeachesAccessibility2018}~\cite{shinoharaWhoTeachesAccessibility2018}.
The core part of this instrument is reproduced below for the reader's benefit:

\begin{enumerate}
  \item How would you rate your knowledge of accessibility: \textsc{Novice}, \textsc{Beginner}, \textsc{Competent}, \textsc{Proficient}, \textsc{Expert}? \emph{[Select One]} \label{qn:a11yKnowledge}
  \item Do you teach courses that incorporate topics about accessibility? \emph{[Yes/No]} \label{qn:teach-a11y}
        \begin{itemize}
          \item If \emph{Yes}, go to the next question; Otherwise, go to Q.~\ref{qn:allteachers}.
        \end{itemize}
  \item Title of your course that incorporated topics about accessibility (If more than one, select one that had the most accessibility content) \emph{[Short text answer]}
  \item Is this an elective course? \emph{[Yes/No]}
  \item This course is typically taken by \textsc{First Year}/\textsc{second year}/
  \textsc{third year}/\textsc{fourth year}/\textsc{Postgraduate} students. \emph{[Multi-select]}
  \item \label{qn:a11y-LOs} What accessibility learning objectives does your course cover? \emph{[Multi-select: 
  (a) Understanding technological barriers faced by PWDs, 
  (b) Understanding design concepts such as universal design, 
  (c) Engaging with individuals from diverse populations appropriately, 
  (d) Evaluating software by accessibility standards such as WCAG,
  (e) Developing accessible web technologies,
  (f) Understanding legal accessibility rights and regulations,
  (g) Understanding different models of disabilities,
  (h) Developing with accessibility-focused tools such as Android Accessibility Suite,
  (i) Employing design techniques: personas, paper prototyping, high-fidelity prototyping
  (j) Other---specify]}
  \item What readings are made available to students? \emph{[Short ans.]}
  \item What pedagogies do you use to teach accessibility? \emph{[Multi-select: (a) Lectures, (b) Active Learning, (c) Collaborative Learning, (d) Cooperative Learning, (e) Blended Learning, (f) Other---specify.]}
  \item In your course, how often do students interact with people with disabilities? \emph{[Select one: (a) More than once a month, (b) Once a month, (c) 2-3 times a semester, (d) Once a semester, (e) Never]}
  \item How many years of experience do you have teaching acces\-sibility-related topics? \emph{[Short answer]}
  \item \label{qn:allteachers}$\triangleright$ What are the barriers to incorporating accessibility topics in your teaching? \emph{[Multi-select]}
         \begin{itemize}
           \item Not a core part of the curriculum
           \item Do not know enough to teach it
           \item Lack of appropriate textbooks
           \item Lack of students and administrator awareness
           \item Lack of support for topics addressing real challenges for people with disabilities
           \item Difficult engaging students
           \item Lack of demand in the industry
           \item Difficult to recruit people with disabilities
         \end{itemize}
  \item To what extent do you agree or disagree with this statement: ``Accessibility should be taught as part of computer science''. \emph{[A 5-point Agree-Disagree Likert scale]}
\end{enumerate}
In the demographics section at the end, in addition to the usual questions about gender and teaching experience, we also asked the participants whether they identify as a PWD or have a PWD in their close circles.

The survey was conducted entirely in English, the primary working language of Indian universities, and underwent a pilot test with faculty members from our institution to ensure clarity and appropriate duration. Following the pilot, only minor adjustments for typos and improved wording were made, confirming the survey's understandability and completion time.

The survey was conducted using tally forms\footnote{\url{https://tally.so/}} with the ability to capture partial responses. Additionally, Google Analytics was configured on the survey link to track the number of unique visits. We launched it by emailing the survey form link to all the 3,689 CS faculty members identified above. However, 194 emails were returned for reasons such as invalid email addresses and full mailboxes. Hence, the total number of emails sent successfully is 3,495. We also requested mailing groups such as ACM India iSIGCSE (India Chapter of Special Interest Group on Computer Science Education) to forward it to their membership. The initial email was sent on February 1, 2024, and the survey remained active until the conclusion of February. A reminder email was subsequently sent two weeks after the initial launch. The landing page of the survey showed the informed consent form approved by the institute's Human Ethical Committee (HEC). 
We received responses from 75 CS faculty and 28 partial responses (More on this in Sec \ref{sec:findings}). 

\subsection{The interviews}
\label{sec:method:interview}

Given the unique context in India described in Sec.~\ref{sec:intro}, beyond replicating the US and Switzerland studies, we wanted to further explore what digital accessibility and teaching accessibility mean to the Indian CS faculty.
We now describe the protocol we designed for one-on-one interviews to investigate these questions, the data collected, and our analysis methods.

\subsubsection{Participants}
Of the 75 survey respondents, 37 expressed willingness to engage in further research.
We sent them an email invitation for the interviews on a single day, followed by reminders after two days.
13 of these 37 participated in the interviews conducted via video conferencing (using Google Meet).

\subsubsection{The Interview Protocol}
The core part of the protocol to address RQs 3 and 4, is shown below:

\begin{enumerate}
  \item Do you think it is important to develop software products that are usable by everyone?
        \begin{itemize}
          \item If yes, why do you think most real-world software falls short of meeting accessibility criteria?
          \item If not, how do you think should a person with certain disabilities, such as visual impairment, interact with the world that is becoming increasingly digitized?
        \end{itemize}
  \item Let's revisit a question from the online survey: ``To what extent do you agree or disagree with the statement: \emph{Accessibility should be taught as part of computer science}?'' Can you answer it on a scale of `Strongly Agree' to `Strongly Disagree'? Can you please explain your choice of answer?
  \item We then explained the four broad categories of accessibility learning outcomes from literature (See Sec.~\ref{sec:relatedwork})---Empathy, Awareness, Technical Knowledge, and Career Opportunities, and asked:\\
        Can you incorporate one or more of these in your teaching? Please explain how.
  \item If appropriate resources are available, would you be willing to incorporate accessibility topics into your teaching?
        \begin{itemize}
          \item If yes, what resources would you need?
          \item If not, can you explain why?
        \end{itemize}
\end{enumerate}


Interview durations spanned 21 to 51 minutes, with an average of 31 minutes.
Audio/Video recordings and the automatically generated transcripts of the interviews were verified, anonymized, and securely stored in cloud-based storage for subsequent analysis.
Cleaned transcripts were then imported into Delve\footnote{\url{https://delvetool.com}}, a web-based qualitative data analysis tool.

The first author performed in-vivo, axial (using the constant comparison method), and selective coding as described by \citeauthor{saldanaCodingManualQualitative2021}~\citep{saldanaCodingManualQualitative2021}.
In the initial coding round, excerpts were labeled with `codes' based on participants' own words using In-Vivo coding, staying true to participants' intent and meaning.
For the second coding round, axial coding was applied to group and abstract codes into categories. Redundant codes were merged and renamed to enhance organization. The resulting categories formed the `axes' around which supporting codes revolved.

In the final round, selective coding was applied to derive a core category, establish relationships with other categories, validate connections, and refine and develop categories as needed. Given the iterative nature of grounded theory techniques, we periodically reviewed interview transcripts, allowing for the scrutiny of codes, categories, and assumptions. 

%% file: findings.tex

\section{Findings}
\label{sec:findings}

We received 75 complete responses and 28 incomplete responses. Additionally, we analyzed the Google Analytics reports and identified that only 137 unique users visited the survey's landing page. Among these 137 survey visitors, 75.18\% (N=103) of them attempted the survey (20.44\% of them partially and 54.74\% of them fully). Table \ref{tab:responseRates} presents the breakdown of survey response details for the total number of emails sent (3495). \vspace{-4pt}

\begin{table}[h]
  \centering
  \caption{Survey Response Details considering N=3495 \label{tab:responseRates}}
  \vspace{-10pt}
  \begin{tabular}{p{3cm}rr}
    \toprule
     Metric                & Count         &  Rate \\ \midrule
     Visitors of survey    & 137       & 3.91\%  \\
     Complete Responses    & 75        & 2.14\%  \\
     Incomplete Responses  & 28        & 0.80\%  \\
     Total Responses       & 103       & 2.94\%  \\  \bottomrule
  \end{tabular}\vspace{-10pt}
\end{table}

Among these partial responses, 11 individuals identified themselves as `Novices' in accessibility and discontinued the survey after completing Question \ref{qn:a11yKnowledge}. Additionally, 10 respondents indicated that they do not teach accessibility (Question \ref{qn:teach-a11y}) and exited the survey, while two of them responded up to halfway through before discontinuing. Furthermore, five participants identified themselves as teaching accessibility but discontinued the survey after the subsequent question, which inquired about the specifics of their accessibility-related course content.

Those who identified as novices in accessibility might have felt overwhelmed by the survey's content, leading them to discontinue their participation. Similarly, individuals who indicated that they do not teach accessibility may have felt that the survey was irrelevant to them and chose to exit. The five respondents affirming their engagement in teaching accessibility may have been initially influenced by social desirability bias but subsequently exited when further questioned about course specifics, or they may have genuinely forgotten the details about the course and decided to quit the survey.

These 28 partial responses were excluded from further analysis because their incomplete responses did not provide sufficient data to understand their views on teaching digital accessibility. Since they did not engage with the relevant sections of the survey, their responses would not accurately reflect their perspectives on the topic. 

The 75 complete responses from CS faculty were from 57 different institutes.
41 out of these 75 are male, 31 are female, and 3 chose not to disclose their gender.
Participants have teaching experience ranging from 1 to 33 years (with a Standard Deviation of 8.77, a mean of 13.3, and a median of 11).

Only 3 of the 75 respondents self-identified as accessibility `Expert,' while 19 others as `Proficient,' 32 as `Competent,' 9 as `Beginners,' and 12 as `Novices.'

Four participants self-identified as individuals with disabilities: one with locomotor disability, two with visual impairment, and one with hearing loss. 48 of the respondents reported that they knew someone with a disability, either as a close family member, a close friend, a personal acquaintance, or a professional acquaintance.

Faculty from a wide range of institutions participated in the study. Table~\ref{tab:participants-nirf} shows the distribution of participants based on their institution's NIRF\footnote{The official framework for ranking higher education institutes across India; see \url{https://www.nirfindia.org/2023/OverallRanking.html}.} ranking.

\vspace{-10pt}
\begin{table}[ht]
  \centering
  \caption{NIRF Ranking of the Participants' Institutes\label{tab:participants-nirf}}
  \vspace{-8pt}
  \begin{tabular}{rrr}
    \toprule
    NIRF Rank           & Survey (75) & Interview (13) \\ \midrule
    1--10               & 6           & 1             \\
    11--25              & 22          & 5             \\
    26--50              & 1           & 0             \\
    51--100             & 16          & 2             \\
    101--200            & 6           & 1             \\
    Others (Not Ranked) & 24          & 4             \\ 
    \bottomrule
  \end{tabular}\vspace{-10pt}
\end{table}

Of the 13 \emph{interview} participants, 3 are males, and 10 are females. Their teaching experience ranges from 1 to 30 years, with a Standard Deviation of 9.54, a mean of 15.59, and a median of 14.28. One participant (P12) taught accessibility in a web development course. 

\subsection{RQ1: Who is teaching accessibility?}
In this section, we delve into the faculty members who indicated their involvement in teaching accessibility. Overall, 10.67\% of faculty (N=8) responded "Yes" to the question, "Do you teach courses that incorporate topics about accessibility?". These faculty represented six unique colleges and universities (with 2 being in the top 25 ranked NIRF institutes). Thus, with the caveat of not knowing the prevalence of faculty who did not respond to the survey, an estimate of the floor of the total number of computing and information science/technology faculty in India teaching accessibility is at least 8 out of 3,495 of the faculty who received the survey, about 0.22\%. 

Experience in teaching accessibility ranged from 1 year to 15 years, with an average of 3.75 years, a median of 2.5, and a standard deviation of 4.6. Regarding self-reported accessibility knowledge, four participants rated themselves as \textsc{`Proficient'}, three as \textsc{`Competent'}, and one as a \textsc{`Beginner'}.

Table \ref{tab:genderData} illustrates a higher proportion of respondents who indicated not teaching accessibility identified as male (55.22\%) in comparison to those identifying as female (40.29\%). This trend aligns with findings from previous studies conducted in the US and Switzerland. In the US study, 75.9\% of male respondents and 19.5\% of female respondents reported not teaching accessibility, while in the Switzerland study, the figures were 82\% for males and 18\% for females. In contrast to the US survey results, where a notable number of respondents teaching accessibility identified as female, our study did not reveal a significant difference ($\chi^2$(d \textit{f}=1, N=72) = 0.177, \textit{p} = 0.67). Similar findings (no significance) were observed in the Switzerland study as well. 

\begin{table}[ht]
  \centering
  \caption{Gender identity and who teaches accessibility \label{tab:genderData}}
  \vspace{-8pt}
  \begin{tabular}{p{3cm}rrr}
    \toprule
                    & Female & Male & Prefer not to say \\ \midrule
    All Respondents &  41.33\% & 54.66\%  & 4.00\% \\
    Who Teach       &  50.00\% & 50.00\%  & 0.00\%    \\
    Who Do Not      &  40.29\% & 55.22\%  & 4.47\%  \\
    \bottomrule
  \end{tabular}\vspace{-8pt}
\end{table}

For participants who indicated teaching accessibility, we inquired about their familiarity with disability, the specific details of accessibility topics covered in their instruction, and the pedagogical approaches employed in their teaching. Of those who reported teaching accessibility, 1 of them reported having a disability themselves (preferred not to disclose the disability), and 6 of them reported knowing a person with a disability in their family, friends, or personal and professional circles, as indicated in Table \ref{tab:KnowAPWD}, Column 3. Note that this was a multi-select question, and the sum need not add up to 100\%. Among the 75 respondents, 64\% knew someone with a disability. However, as shown in Table \ref{tab:KnowAPWD}, those who taught accessibility were significantly more likely to report knowing someone with a disability ($\chi^2$(d \textit{f}=1, N=75) = 29.47, \textit{p} = 0.00001). This finding is consistent with the findings of the other two studies. \vspace{-4pt}

\begin{table}[H]
  \centering
  \caption{Proportion of faculty who do, do not teach accessibility and who know someone with a disability \label{tab:KnowAPWD}}
  \vspace{-8pt}
  \begin{tabular}{p{4cm}rrr}
    \toprule
    Relationship                    & All & Teaches & Doesn't \\ \midrule
    No                              & 36.00\% & 25.00\% & 37.31\% \\
    Yes,close family                & 21.33\% & 37.50\% & 19.40\% \\
    Yes,close friends               & 13.33\% & 25.00\% & 11.94\%  \\
    Yes,personal acquaintances      & 12.00\% & 12.50\%  & 11.94\%  \\
    Yes,professional acquaintances  & 33.33\% & 25.00\% & 34.32\%  \\
    \bottomrule
  \end{tabular}\vspace{-8pt}
\end{table}

Out of the 75 responses received, 62 participants provided details about their research interests and areas of expertise. Each unique specialization mentioned was counted, considering many respondents listed multiple areas of expertise. These specializations were then classified according to the 2023 ACM Curriculum "Knowledge Areas,"\footnote{\url{https://csed.acm.org/knowledge-areas/}} with minor variations in terminology. Following this categorization, the top five areas of expertise identified across all respondents were Artificial Intelligence (AI), Software Engineering (SE), Algorithmic Foundations (AL), Networking and Communication (NC), and Graphics and Interactive Techniques (GIT). Notably, four of these five areas align with those reported in the US and Switzerland studies as their top categories. 

Among the eight respondents who reported teaching accessibility and provided details about their expertise, only one specialized in human-computer interaction (HCI), while the others possessed expertise in various fields such as software engineering, AI, information systems, and eGovernance. This contrasts with the findings from the US and Switzerland surveys, where most faculty teaching accessibility typically had expertise in human-computer interaction/ human and social factors in computing.

We asked respondents who teach accessibility to indicate which objectives they focus on in their teaching and the accessibility learning Objectives (LO). Table \ref{tab:A11yLO} compares the responses to the relevant question from all three studies. The rows are organized in descending order of responses on the US survey to match the corresponding table in the original paper.

\begin{table}[H]
  \centering
  \caption{Learning Objectives Faculty reported teaching a11y\label{tab:A11yLO}}
  \vspace{-10pt}
  \begin{tabular}{p{4.8cm}rrr}
    \toprule
    Learning Objective            & US      & Switz. & India \\\midrule
    Understand technology barriers faced by people with disabilities  & 66.1\pc & 50.00\pc  & 62.50\pc \\
    Understand design concepts: universal design, ability-based design, inclusive design, participatory design, etc. & 65.90\pc & 37.50\pc & 50.00\pc \\
    Engage with individuals from diverse populations appropriately & 40.00\pc & 12.50\pc & 37.50\pc \\
    Evaluate web pages by accessibility standards and heuristics (e.g., W3C, WCAG) & 36.5\pc & 62.50\pc & 75.00\pc \\
    Be able to develop accessible web technologies (e.g., use of alt-tags, captioning videos, and describing images) & 36.00\pc & 25.00\pc & 62.50\pc \\
    Be able to employ design techniques: personas, paper prototyping, high-fidelity prototyping & 35.20\pc & 25.00\pc & 0.00\pc \\
    Understand legal accessibility regulations & 31.50\pc & 25.00\pc & 12.50\pc \\
    Understand the different models of disability (e.g., social, medical or legal models) & 15.2\pc & 12.50\pc & 25.00\pc \\
    Be able to develop with accessibility-focused technical languages and tools (Apple’s UI Accessibility Programming Interface, Android’s Accessibility Events, Universal Windows Platform) & 6.10\pc & 0.00\pc & 12.50\pc \\
    Other & 4.80\pc & 0.00\pc & 0.00\pc \\
    None of the above & 3.20\pc & 0.00\pc & 0.00\pc \\
    \bottomrule
  \end{tabular}\vspace{-10pt}
\end{table}

\begin{table*}[t]
  \centering
  \caption{Details of the Course teaching accessibility \label{tab:CourseDetails}}
  \vspace{-8pt}
  \begin{tabular}{p{4.5cm}rrrr}
    \toprule
    Course Title & Course Type & Course Offered in & Reading Resources offered & Students interacted with PWD \\ \midrule
    User interface design & Elective & 4th Year & Web resources & Never \\
    Human-Computer Interface & Elective & 3rd Year & Books and econtent & Once a semester \\
    Deep Thinking & Elective & 4th Year & Journals and Papers & More than once a month \\
    Software Engineering and Testing & Mandatory & 2nd/3rd Year & Books and eBooks & More than once a month \\
    Web Technology and Multimedia & Mandatory & 3rd Year & Web resources & Never \\
    Web development & Mandatory & Postgraduate & WCAG Guidelines & Never \\
    Human-Computer Interface & Elective & 3rd Year & Textbook & Never \\
    An Introduction to Accessibility in the Global South & Elective & 4th Year & Articles on Accessibility & Once a Month \\
    \bottomrule 
  \end{tabular}\vspace{-10pt}
\end{table*}

In addition to replicating previous studies, we delved deeper into how accessibility is taught by surveying participants who integrated it into their courses. We asked about the courses in which they integrated accessibility, whether the course was offered as an elective, the targeted year of students, the reading materials provided, and whether students engaged with individuals with disability as part of the course. The findings are summarized in Table \ref{tab:CourseDetails}.

We also asked about the pedagogy used to teach accessibility. All of them used lectures and class meetings, 75\% of them used blended learning such as flipped classrooms, and 50\% used active learning involving in-class activities. 37.5\% also mentioned using collaborative learning involving peer instruction. Again, the top methods are consistent with the findings from the other two studies.

\subsection{RQ2: Barriers to teaching accessibility}

Like the US and Switzerland studies, our second research question investigates factors CS educators perceive as barriers to teaching accessibility.
Table~\ref{tab:barriers} compares the responses to the relevant question from all three studies. 
The rows are organized in descending order of responses on the US survey to match the corresponding table in the original paper. Irrespective of the actual numbers, the top four barriers are the same across all three studies.

\vspace{-8pt}

\begin{table}[H]
  \centering
  \caption{Barriers to teaching accessibility\label{tab:barriers}}
  \vspace{-10pt}
  \begin{tabular}{p{4cm}rrr}
    \toprule
                                  & US      & Switz. & India \\\midrule
    Not a core part of curriculum & 52.30\pc & 77.00\pc       & 70.66\pc \\
    Don't know enough to teach it & 49.10\pc & 21.00\pc       & 50.66\pc \\
    Lack of appropriate textbook  & 14.90\pc & 9.00\pc        & 40.00\pc \\
    Lack of students and administrator awareness & 14.10\pc & 9.00\pc        & 36.00\pc \\
    None of the above             & 6.00\pc  & 0.00\pc        & 0.00\pc  \\
    Other                         & 13.10\pc & 4.00\pc        & 1.33\pc  \\
    Lack of support for topics addressing real challenges for people with disabilities                  & 13.10\pc & 0.00\pc        & 29.33\pc \\
    Difficult engaging students   & 10.20\pc & 6.00\pc        & 13.33\pc \\
    Lack of demand in industry    & 8.20\pc  & 4.00\pc        & 14.66\pc \\
    Difficult to recruit people with disabilities & 7.20\pc  & 0.00\pc        & 17.33\pc \\
    All of the above              & 6.00\pc  & 0.00\pc        & 0.00\pc  \\\bottomrule
  \end{tabular}\vspace{-10pt}
\end{table}

Similar to the other two studies, many respondents `Agree' or `Strongly Agree' with the statement `Accessibility should be taught as part of computer science' (See Table~\ref{tab:a11y-in-cs}). In our study, more than 80\% of the respondents either `Agree' or `Strongly Agree' that accessibility should be taught as part of CS. 

\begin{table}[H]
  \centering
  \caption{Accessibility should be taught as part of CS\label{tab:a11y-in-cs}}\vspace{-10pt}
  \begin{tabular}{p{4cm}rrr}
    \toprule
                               & US   & Switz. & India \\\midrule
    Strongly Agree             & 4.1\pc  & 25.0\pc        & 38.66\pc \\
    Agree                      & 42.6\pc & 38.0\pc        & 44.00\pc \\
    Neither Agree nor Disagree & 29.9\pc & 31.0 \pc       & 12.00\pc  \\
    Disgree                    & 4.8\pc  & 5.0\pc         & 2.66\pc  \\
    Strongly Disagree          & 4.1\pc  & 0.0 \pc        & 2.66\pc  \\\bottomrule
  \end{tabular}
\end{table}

\subsection{RQ3: What are the faculty members' perceptions about accessibility?}

Given over a quarter of the respondents considered themselves Beginners or Novices in digital accessibility, we explored their perceptions of accessibility through the interviews.

\begin{figure*}[ht]
  \centering
  \frame{\includegraphics[width=2.2\columnwidth]{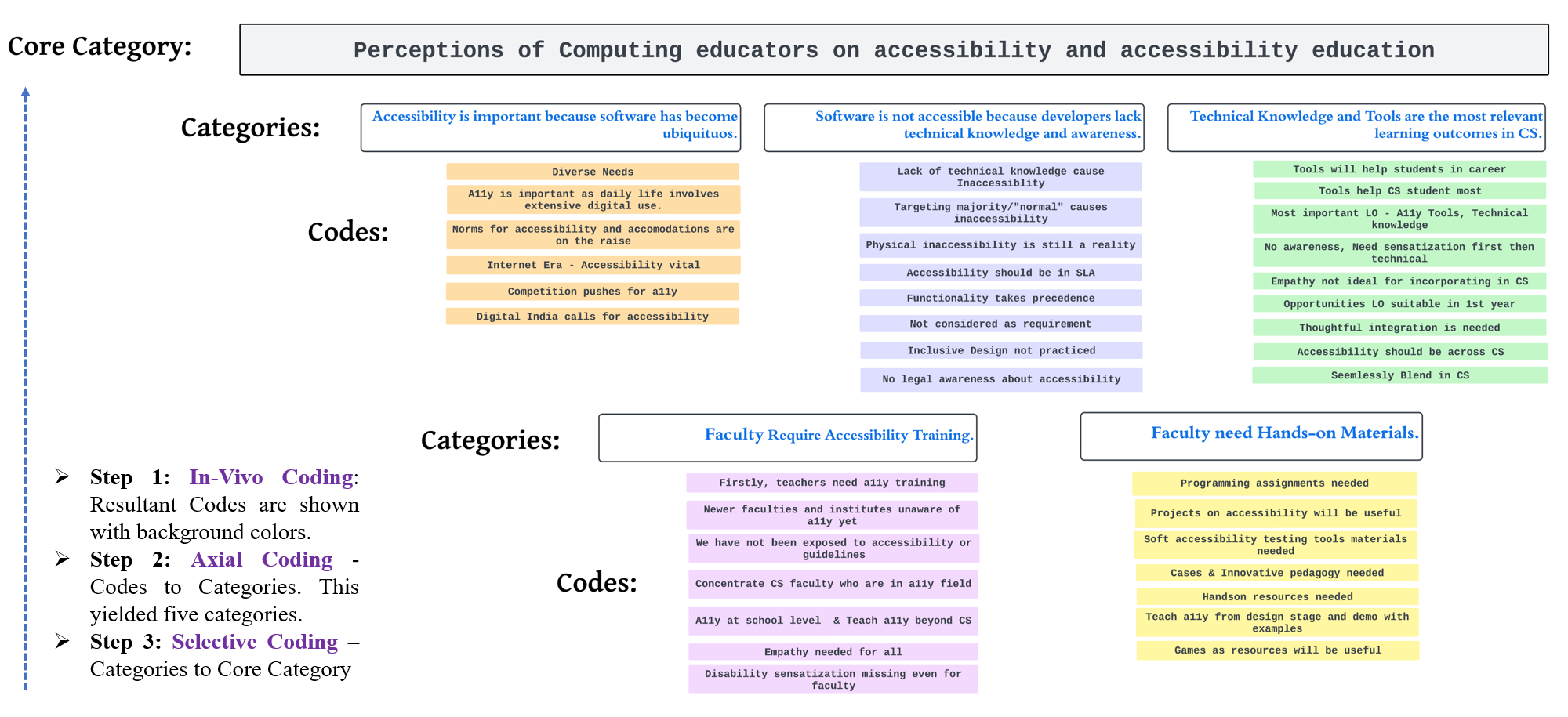}}\vspace{-8pt}
  \caption{Codes and Categories Obtained from Qualitative Analysis }
  \label{fig:conceptmap}\vspace{-8pt}
\end{figure*}

The final in-vivo codes, categories, and core category found in our work are represented in Fig \ref{fig:conceptmap}. We now present the code categories that emerged after analyzing the transcripts (as described in Sec.~\ref{sec:method:interview}) relevant to answer RQ3.  

\subsubsection{\textsc{Accessibility is important because software has become ubiquitous}}

Participants stated that it is essential to develop accessible software because ``We are living in a kind of Internet era.'' (P7), and ``[P]eople are so well versed with the usage of the software products\ldots their needs have grown'' (P1).

One of the respondents mentioned a particular requirement for software to be usable for the elderly: ``\ldots for elderly people like the font size should be bigger and stuff like that.'' (P9)

Another respondent (P2) related their experience of using software for several daily operations like paying bills and wondered how difficult it would be for someone with disabilities to perform those operations if the software is not accessible.

P2 also mentioned that other aspects, like accessibility in physical infrastructure, are also becoming important and visible: ``I think [the] government has made some changes and because of which, at least in the new constructions, these things [such as wheelchair ramps] are being taken care of.''

 Another respondent mentioned an increase in sensitivity due to government regulations:
 \begin{quote}
   [PWDs] are put up in classrooms which are on the ground floor or something, so they don't have to climb the stairs. So I think the sensitivity \ldots even in that physical access itself, perhaps was not there, and it has started happening now with some mandates which government has put up. (P4)
 \end{quote}

Most participants recognized the ongoing digital transformation worldwide, with one participant (P10) emphasizing, " ... Digital India calls for accessibility". Digital India is an initiative introduced by the Indian government to enhance citizens' access to government services through improved online infrastructure and expanded Internet connectivity.

Nine out of the thirteen participants' responses contributed to this category.

\subsubsection{\textsc{Software is not accessible because developers lack technical knowledge and awareness}}

Most responses to the question `Why do you think most real-world software falls short of meeting accessibility criteria?' explicitly specified a lack of awareness and technical knowledge. Ten respondents' data contributed to this category.

For instance, a respondent mentioned: 

\begin{quote}
    \ldots\ lack of awareness, people who are designing need to have an idea about accessibility ... not only in terms of disability but if you evaluate an application, most of them are created based on the experience they [developers] are having ... inclusive design is not practiced. (P12) 
\end{quote}

The same respondent also brought up the topic of accessibility not being enforced as strictly as it should be by mentioning: 

\begin{quote}
\ldots\ conformance to standards ...we have guidelines, standards, W3C publishes these but all these are not enforced in a stricter basis, whereas when it comes to others like data [privacy] it is more strict. No one takes the accessibility reports seriously, whereas if it's security [bug] or if I am not using https, the application is blocked ... not enforcing accessibility seriously is [contributing] to inaccessibility. (P12)
\end{quote}
Another respondent stated that since the majority of the users do not require accessibility features, they are not implemented:

\begin{quote}
  \ldots\
  if I need to spend 90\% of my time in getting that 10\%, then why to do that. (P2)
\end{quote}

P6 voices a similar opinion:

 \begin{quote}
   [Software] is mainly developed keeping in mind the number. I mean, who will be the largest number of users? \ldots I guess that the \emph{normal} users are more, so that could be one of the reasons. (Emphasis added.)
 \end{quote}

Such responses suggest the CS faculty require sensitization training as well. 

\subsection{RQ4. What kind of support do faculty seek to teach accessibility in their courses?}



Using the same qualitative coding methods, we analyzed responses to questions about resources the respondents believed they need to incorporate accessibility in their teaching. The following three categories emerged from the coding.


\subsubsection{\textsc{Technical Knowledge is the most relevant learning outcome}}

Most participants (11 out of 13) indicated that awareness about how PWDs interact with software and technical knowledge of implementing and testing accessibility are the most suitable learning outcomes (LO) for CS courses.

Some faculty also felt that empathy is required but opined that it is not an appropriate learning outcome for \emph{CS courses}: ``Like the empathy and all like, I don't think computer programming can teach empathy. We struggle [with] teaching programming fundamentals\ldots, and it's a very cutthroat competition where I don't think empathy would be grasped'' (P9).

P8 suggested accessibility should be introduced when discussing UI/UX design in CS courses.

Surprisingly, only one participant, P9, thought `career opportunities related to accessibility' is a suitable LO but added that it should only be covered in the first year.

\subsubsection{\textsc{Faculty Require Accessibility Training}}

Participants acknowledged the need to learn about accessibility themselves: ``As a computer science teacher\ldots\ I [should] also know what should be taught'' (P7). 

A similar opinion was shared by another participant (P13):

\begin{quote}
    \ldots the use of alternate text for [visually impaired] people or blind people is not known to many of them. Even I was not knowing this because of this it doesn't get conveyed to students. We [teachers] should know the guidelines first \ldots.
\end{quote}

Eight participants contributed to this category. Some participants were concerned about their use of language: 
\begin{quote}
  So like we never had a formal [training] of such a thing, right? So like, I'm not sure, like, so what are the correct things to say to describe a certain individual, right? (P9).
\end{quote}
This emphasizes our point about sensitization training mentioned earlier.

\subsubsection{\textsc{Faculty need hands-on resources}}

Many participants expressed the need to make programming assignments with accessibility components available to incorporate into their courses readily. 

P9 requested modules that teach accessibility topics, especially awareness, through various courses starting from the first year, while P7 suggested making such modules available for courses in all four years.
Some (P4, P5, P9,P10) went one step further and suggested accessibility should be taught from the school level itself; however, they appear to talk about disabilities-sensitization training rather than teaching accessibility in computing courses. P5 requested material, including case studies around accessibility, to introduce accessibility.

A participant mentioned they require materials on guidelines and various accessibility evaluation tools. 

\begin{quote}
    First of all, the guidelines and how to do these [evaluations] all that and if there any tools that can look at an existing product and tell how much accessibility is done and what can be done to improve the accessibility score. \ldots students will also interested and we can also teach it in a more interactive way P(12).
\end{quote}

Another participant emphasized the benefits of game-based learning and requested materials and games to introduce accessibility. P11 mentioned, 

\begin{quote}
    first is material, because we are not capable or we don't know [about accessibility]. Laws and all we can read from the internet but I don't want to introduce this as a theoretical subject but would want to have it like a game. \ldots with a brainstorming session with you or likeminded folks, can come up with 4-5 games like an ice breaker to teach [accessibility] and for them [students], it is fun and thought provoking (P11).
\end{quote}

While some participants identified courses such as Web development, Software Engineering, and Software Testing as good candidates for teaching accessibility (P5, P6, P13, P11), it should be noted that some faculty felt that not all CS courses could integrate accessibility. For instance, P2 stated:
 \begin{quote}
  ``The subjects that I teach [\ldots] focus not much on, you know, the UI part, [but] more on the algorithms and thinking, and stuff like that. So the scope for integration [of accessibility] is much low [in these subjects].
   But I mean, [in some other CS courses,] I foresee the integration to be much more natural\ldots\ I might have to, you know, think this through a bit more, because [\ldots] I mean, there are many places where these things are required.''
 \end{quote}
Some faculty were unsure what resources they needed to integrate accessibility into their courses (P4, P7), but they acknowledged the need for such material from more experienced faculty. 

%% file: discussion.tex

\section{Discussion}
\label{sec:discussion}
In this section, we discuss our findings and highlight both the commonalities and distinctions observed in comparison to the previous two studies. Most of our findings corroborate the findings of the earlier studies. 

Our findings, echoing those of the two preceding studies, can be summarized as follows:
\begin{itemize}
    \item The presence of faculty teaching accessibility is limited, with 20\% of respondents teaching it in the US study, 14.2\% in the Switzerland study, and 10.66\% in our study. When considering the total number of faculty who received the survey, the figures are even lower: only 2.5\% in the US, 11.59\% (due to the smaller email distribution in the Swiss study), and 0.22\% in our study. 
    \item Across the three studies, a greater proportion of respondents who stated they do not teach accessibility identified as male. Those who taught accessibility were likelier to report knowing someone with a disability.
    \item In Switzerland and India, the primary learning objective in teaching accesibility is to evaluate web pages based on accessibility standards and heuristics. Conversely, in the US, the focus is on helping students comprehend the technological barriers experienced by individuals with disabilities. 
    \item In all three studies, respondents advocate for integrating accessibility into CS degrees. However, they identify consistent barriers: the absence of accessibility in the core curriculum, inadequate expertise among CS faculty, and a scarcity of appropriate textbooks.
\end{itemize}

Our findings, contrasting those of the two preceding studies, are as follows:
\begin{itemize}
    \item In the US survey, human-computer interaction (HCI) is the most common expertise (28\%) among those who teach accessibility, and only 1.8\% of HCI experts do not teach accessibility. In the Switzerland survey, half of those who teach accessibility self-reported to be HCI experts. In our survey, however, only one of the respondents reported that HCI is their field of expertise and teaches accessibility.
    \begin{itemize}
        \item Another important finding is the lack of HCI experts among the respondents. We investigated this further by reviewing the websites of CS faculty in the Top 10 institutes ranked by NIRF. Only 9 out of the 420 CS faculty work in HCI (2.14\%) in three of these 10 institutes. 
        \item While there could be several other reasons why eight of these nine faculty did not participate in our survey, it could also be because they do not teach accessibility. As an anecdotal example, the only HCI expert at our institute does not include accessibility in his lessons, and he did not participate in our survey. This is in sharp contrast with the findings of the US and Switzerland.
    \end{itemize}
    \item In our study, none of the faculty teaching accessibility identified themselves as an \textsc{`Expert'}, and three of those who did identify as experts do not teach accessibility. 
    \begin{itemize}
        \item Since none of these self-identified experts participated in the interview, we could not understand why they chose not to teach accessibility despite their expertise. This discrepancy may also indicate a form of self-enhancement or self-boasting bias.
        \item The absence of self-identification as experts among faculty teaching accessibility, as affirmed by one interview participant who teaches accessibility, underscores the pressing need to enhance the accessibility skills of CS faculty. One potential approach could be to collaborate with experts from Western institutions. As emphasized by \citeauthor{agboComputingEducationResearch2023}~\citep{agboComputingEducationResearch2023}, donor institutions, particularly in the West, should prioritize and support higher education initiatives in the Global South as part of the global development agenda. Similarly, researchers from Western countries can benefit from learning from their counterparts in the Global South, gaining insights into how diverse contexts engage with and respond to efforts to promote inclusivity, accessibility, and diversity in computing education.
    \end{itemize}
    \item Additionally, our findings differ from those of the US and Switzerland studies regarding certain barriers to teaching accessibility. While the lack of support for topics addressing real challenges for people with disabilities and difficulty in recruiting people with disabilities were not major barriers in the US and Switzerland, our study revealed that a considerable number of respondents felt these were significant barriers. 
    \begin{itemize}
        \item Our qualitative analysis also underscores the necessity for support in teaching accessibility among CS faculty in India. Faculty in India require assistance not only with instructing accessibility but also with enhancing awareness about disability and accessibility overall, as well as implementing policies to facilitate the recruitment of individuals with disabilities for instructional roles.
    \end{itemize} \vspace{-8pt}
\end{itemize}

As shown in Table \ref{tab:CourseDetails}, five out of eight respondents teaching accessibility do so in an elective course, which may not be pursued by all CS students. While this marks a positive beginning, it's not comprehensive enough. To truly make an impact, accessibility needs to be integrated into core computer science courses. As emphasized by \citeauthor{elglalyHCINeedAccessibility2024a}~\citep{elglalyHCINeedAccessibility2024a}, accessibility should be incorporated as a fundamental aspect of computing education, spanning across different CS disciplines. 

Furthermore, as depicted in Table \ref{tab:CourseDetails}, half of the instructors who teach accessibility fail to involve their students in interactions with individuals with disabilities. According to \citeauthor{connUnderstandingMotivationsFinalyear2020}~\citep{connUnderstandingMotivationsFinalyear2020}, students who engage with persons with disabilities tend to retain accessibility knowledge better over the long term. However, our study, detailed in Table \ref{tab:barriers} and supported by qualitative findings, reveals that recruiting individuals with disabilities for in-class instruction or sensitization sessions on assistive technology poses a significant challenge. This barrier warrants further investigation to uncover its root causes, whether they stem from institutional policies, a lack of connection with people with disabilities, or other factors 

It is worth noting that none of the participants in the interview were aware of the fact that ACM/IEEE CS2013 Curricula~\citep{acmcomputingcurriculataskforceComputerScienceCurricula2013} and the 2023 revisions ~\citep{10.1145/3664191} include accessibility as Core topics, even though it has been in the public domain for a few years.
This points to a larger issue of relatively low participation of the Global South in Computing Education Research (CER).
In a recent literature review, \citeauthor{agboComputingEducationResearch2023}~\citep{agboComputingEducationResearch2023} expounded to this gap: only 7.7\% of CER publications originate in the Global South.

\section{Limitations and Future Work}
\label{sec:limitations}

Our survey has several limitations. We sent out the survey in Feb 2024, when some faculty might have been busy with the start of the semester activities, while a few others might not have been back on campus yet. Additionally, we refrained from using third-party emailing services due to the possibility of institutions blocking inward email access from generic email services. Instead, we utilized personal emails to send invitations, which may have led to response bias as faculty members who recognized the survey authors might have been more inclined to respond.

Despite the email invitation stating that no prior knowledge of accessibility was necessary, some participants may have considered the survey topic irrelevant to their work or interests. This is indicated by the low visit rate (to the survey link) of only 3.96\%, as observed in the Google Analytics results. Several factors could contribute to this lack of engagement, such as competing priorities, time constraints, or a perceived mismatch with their research or teaching focus.

The 2.66\% of faculty members who strongly disagreed with integrating accessibility into computer science did not participate in the interview process. As a result, we could not gain insights into their reasons for holding this belief. However, none of them was, nor did they report knowing, a \pwd\ in their close circles.

Despite such limitations, some findings are clear:\vspace{-2pt}
\begin{itemize}
  \item Very few CS faculty in India are teaching accessibility.
  \item There are few HCI faculty in CS departments in India, and, unlike in the West, they do not seem to teach accessibility topics in CS courses. (It is likely that accessibility is taught in Design disciplines, though. However, this was beyond the scope of our study and we focus on CS Education).
  \item However, many faculty in India recognize the need to teach these topics in computing courses, and the barriers they foresee are similar to the ones faculty in other parts of the world experience.
  \item They feel they need training and hands-on resources they can readily incorporate into their courses to bridge the gap.
\end{itemize}

We have some recommendations to address this situation:
\begin{itemize}
  \item \textbf{Developing a Community of Practitioners} --- 62 of the 75 survey participants (and all thirteen interview participants) favor integrating accessibility into CS curricula. They represent colleges and universities with diverse administrative challenges and student populations. We recommend developing a community of CS faculty interested in accessibility so they can learn from each other, get support, and create a more meaningful impact. Such a community would serve as a platform for educators, researchers, industry professionals, and individuals passionate about accessibility to collaborate and collectively work towards advancing accessibility initiatives. While initiatives such as Teach Access~\citep{waltherBroadeningParticipationTeaching2021}, and AccessComputing\cite{10.1145/2968453} are already leading such efforts in the US, considering the needs of the Global South (language barriers, cultural differences, powers or lack thereof of a typical college instructor), perhaps a more \emph{home grown} effort is necessary.
  \item \textbf{Hands-on Teaching Resources} --- Creating practical teaching materials that seamlessly integrate into existing computer science courses, allowing faculty to incorporate accessibility content into their regular curriculum effortlessly. This involves adapting resources from various researchers, as documented in the comprehensive repository `Including Accessibility In Computer Science'\footnote{\url{https://accessibilityeducation.github.io/index.html}}, to cater to the diverse educational requirements and contexts within the Indian academic system.
  \item \textbf{Sensatization Workshops} --- As emphasized by many interview participants, there exists a notable deficiency in awareness regarding accessibility among educators, administrators, curricula designers, and students in India. Addressing this gap requires the implementation of sensitization workshops nationwide, possibly integrated into conferences or specialized capacity-building events for faculty. These workshops could encompass a range of topics, including disability awareness, assistive technology, and accessibility practices, focusing on incorporating relevant use cases from the Global South to enhance engagement and relevance. Furthermore, as participants suggested, there is potential for integrating discussions on disability etiquette, introduction to accessibility, and assistive technology into K-12 computing curricula in India. By introducing these concepts at a younger age, students can develop a deeper understanding of the importance of accessibility and foster a more inclusive mindset from the outset.
  \item \textbf{Engaging Industry Professionals} --- In the meantime, it's essential to acknowledge the importance of mutual learning between academia and industry in accessibility education, as underscored by \citeauthor{coverdaleTeachingAccessibilityShared2022}~\citep{coverdaleTeachingAccessibilityShared2022}. This collaborative approach promotes the exchange of knowledge, experiences, and best practices, benefiting both sectors. India has notably introduced the concept of a Professor of Practice (PoP)\footnote{\url{https://pop.ugc.ac.in/}}, which allows accomplished industry leaders to participate in teaching activities within academic institutions actively. The engagement of accessibility professionals in teaching endeavors helps students develop a thorough understanding of accessibility principles, practices, and their practical applications across various contexts.
\end{itemize} 

Strategic collaborations on these can lead to a significant shift in accessibility education in India and the Global South, scaling up the efforts to teach accessibility worldwide.

In conclusion, this replication study corroborates the earlier surveys' key findings in the Indian context: very few CS faculty teach accessibility, and the top barriers to teaching accessibility they perceive are common across the world. Furthermore, it highlights the need for accessibility training, disability sensitization, and exposure to ongoing CS education research in the Global South.

%% file: main.bbl
\begin{thebibliography}{29}

  
  \ifx \showCODEN    \undefined \def \showCODEN     #1{\unskip}     \fi
  \ifx \showDOI      \undefined \def \showDOI       #1{#1}\fi
  \ifx \showISBNx    \undefined \def \showISBNx     #1{\unskip}     \fi
  \ifx \showISBNxiii \undefined \def \showISBNxiii  #1{\unskip}     \fi
  \ifx \showISSN     \undefined \def \showISSN      #1{\unskip}     \fi
  \ifx \showLCCN     \undefined \def \showLCCN      #1{\unskip}     \fi
  \ifx \shownote     \undefined \def \shownote      #1{#1}          \fi
  \ifx \showarticletitle \undefined \def \showarticletitle #1{#1}   \fi
  \ifx \showURL      \undefined \def \showURL       {\relax}        \fi
  \providecommand\bibfield[2]{#2}
  \providecommand\bibinfo[2]{#2}
  \providecommand\natexlab[1]{#1}
  \providecommand\showeprint[2][]{arXiv:#2}
  
  \bibitem[Sta(2022)]%
          {StackOverflow20222022}
   \bibinfo{year}{2022}\natexlab{}.
  \newblock \bibinfo{title}{Stack {{Overflow}}: 2022 {{Developer Survey}}}.
  \newblock \bibinfo{howpublished}{https://survey.stackoverflow.co/2022}.
  \newblock
  
  
  \bibitem[UGC(2022)]%
          {UGCAnnualReport2022}
   \bibinfo{year}{2022}\natexlab{}.
  \newblock \bibinfo{booktitle}{\emph{{{UGC Annual Report}} 2021-22}}.
  \newblock \bibinfo{type}{{T}echnical {R}eport}. \bibinfo{institution}{{University Grants Commission}}, \bibinfo{address}{{New Delhi, India}}.
  \newblock
  
  
  \bibitem[Bri(2023)]%
          {BridgingAccessibleTechnology2023}
   \bibinfo{year}{2023}\natexlab{}.
  \newblock \bibinfo{title}{Bridging {{The Accessible Technology Skills Gap}}: {{Teach Access}} 2023 {{Survey}} \& {{Analysis}}}.
  \newblock \bibinfo{howpublished}{https://teachaccess.org/accessibility-skills-gap/}.
  \newblock
  
  
  \bibitem[{ACM Computing Curricula Task Force}(2013)]%
          {acmcomputingcurriculataskforceComputerScienceCurricula2013}
  \bibfield{author}{\bibinfo{person}{{ACM Computing Curricula Task Force}}.} \bibinfo{year}{2013}\natexlab{}.
  \newblock \bibinfo{booktitle}{\emph{Computer {{Science Curricula}} 2013: {{Curriculum Guidelines}} for {{Undergraduate Degree Programs}} in {{Computer Science}}}}.
  \newblock \bibinfo{publisher}{{ACM, Inc}}.
  \newblock
  \showISBNx{978-1-4503-2309-3}
  \urldef\tempurl%
  \url{https://doi.org/10.1145/2534860}
  \showDOI{\tempurl}
  
  
  \bibitem[Agbo et~al\mbox{.}(2023)]%
          {agboComputingEducationResearch2023}
  \bibfield{author}{\bibinfo{person}{Friday~Joseph Agbo}, \bibinfo{person}{Maria Ntinda}, \bibinfo{person}{Sonsoles {L{\'o}pez-Pernas}}, \bibinfo{person}{Mohammed Saqr}, {and} \bibinfo{person}{Mikko Apiola}.} \bibinfo{year}{2023}\natexlab{}.
  \newblock \showarticletitle{Computing {{Education Research}} in the {{Global South}}}.
  \newblock In \bibinfo{booktitle}{\emph{Past, {{Present}} and {{Future}} of {{Computing Education Research}} : {{A Global Perspective}}}}, \bibfield{editor}{\bibinfo{person}{Mikko Apiola}, \bibinfo{person}{Sonsoles {L{\'o}pez-Pernas}}, {and} \bibinfo{person}{Mohammed Saqr}} (Eds.). \bibinfo{publisher}{{Springer International Publishing}}, \bibinfo{address}{{Cham}}, \bibinfo{pages}{311--333}.
  \newblock
  \showISBNx{978-3-031-25336-2}
  \urldef\tempurl%
  \url{https://doi.org/10.1007/978-3-031-25336-2_15}
  \showDOI{\tempurl}
  
  
  \bibitem[Baker et~al\mbox{.}(2020)]%
          {bakerSystematicAnalysisAccessibility2020}
  \bibfield{author}{\bibinfo{person}{Catherine~M. Baker}, \bibinfo{person}{Yasmine~N. {El-Glaly}}, {and} \bibinfo{person}{Kristen Shinohara}.} \bibinfo{year}{2020}\natexlab{}.
  \newblock \showarticletitle{A Systematic Analysis of Accessibility in Computing Education Research}. In \bibinfo{booktitle}{\emph{Proceedings of the 51st {{ACM}} Technical Symposium on Computer Science Education}} \emph{(\bibinfo{series}{{{SIGCSE}} '20})}. \bibinfo{publisher}{{Association for Computing Machinery}}, \bibinfo{address}{{New York, NY, USA}}, \bibinfo{pages}{107--113}.
  \newblock
  \showISBNx{978-1-4503-6793-6}
  \urldef\tempurl%
  \url{https://doi.org/10.1145/3328778.3366843}
  \showDOI{\tempurl}
  
  
  \bibitem[Bhatia et~al\mbox{.}(2023)]%
          {bhatiaIntegratingAccessibilityMobile2023}
  \bibfield{author}{\bibinfo{person}{Jaskaran~Singh Bhatia}, \bibinfo{person}{Parthasarathy P~D}, \bibinfo{person}{Snigdha Tiwari}, \bibinfo{person}{Dhruv Nagpal}, {and} \bibinfo{person}{Swaroop Joshi}.} \bibinfo{year}{2023}\natexlab{}.
  \newblock \showarticletitle{Integrating {{Accessibility}} in a {{Mobile App Development Course}}}. In \bibinfo{booktitle}{\emph{Proceedings of the 54th {{ACM Technical Symposium}} on {{Computer Science Education V}}. 1}}. \bibinfo{publisher}{{ACM}}, \bibinfo{address}{{Toronto ON Canada}}, \bibinfo{pages}{1021--1027}.
  \newblock
  \showISBNx{978-1-4503-9431-4}
  \urldef\tempurl%
  \url{https://doi.org/10.1145/3545945.3569825}
  \showDOI{\tempurl}
  
  
  \bibitem[Cohen et~al\mbox{.}(2005)]%
          {cohenAccessibilityIntroductoryComputer2005}
  \bibfield{author}{\bibinfo{person}{Robert~F. Cohen}, \bibinfo{person}{Alexander~V. Fairley}, \bibinfo{person}{David Gerry}, {and} \bibinfo{person}{Gustavo~R. Lima}.} \bibinfo{year}{2005}\natexlab{}.
  \newblock \showarticletitle{Accessibility in {{Introductory Computer Science}}}. In \bibinfo{booktitle}{\emph{Proceedings of the 36th {{SIGCSE Technical Symposium}} on {{Computer Science Education}}}} \emph{(\bibinfo{series}{{{SIGCSE}} '05})}. \bibinfo{publisher}{{ACM}}, \bibinfo{address}{{New York, NY, USA}}, \bibinfo{pages}{17--21}.
  \newblock
  \showISBNx{978-1-58113-997-6}
  \urldef\tempurl%
  \url{https://doi.org/10.1145/1047344.1047367}
  \showDOI{\tempurl}
  
  
  \bibitem[{Department of Higher Education, Government of India}(2021)]%
          {departmentofhighereducationgovernmentofindiaAllIndiaSurvey2021}
  \bibfield{author}{\bibinfo{person}{{Department of Higher Education, Government of India}}.} \bibinfo{year}{2021}\natexlab{}.
  \newblock \bibinfo{title}{All {{India}} Survey on {{Higher Education}} 2020-21}.
  \newblock
  \newblock
  
  
  \bibitem[{El-Glaly}(2020)]%
          {el-glalyTeachingAccessibilitySoftware2020}
  \bibfield{author}{\bibinfo{person}{Yasmine~N. {El-Glaly}}.} \bibinfo{year}{2020}\natexlab{}.
  \newblock \showarticletitle{Teaching Accessibility to Software Engineering Students}. In \bibinfo{booktitle}{\emph{Annual Conference on Innovation and Technology in Computer Science Education, {{ITiCSE}}}} \emph{(\bibinfo{series}{{{SIGCSE}} '20})}. \bibinfo{publisher}{{Association for Computing Machinery}}, \bibinfo{address}{{Portland, OR, USA}}, \bibinfo{pages}{121--127}.
  \newblock
  \showISSN{1942647X}
  \urldef\tempurl%
  \url{https://doi.org/10.1145/3328778.3366914}
  \showDOI{\tempurl}
  
  
  \bibitem[Freire et~al\mbox{.}(2013)]%
          {freireAccessibilityWebMultimedia2013}
  \bibfield{author}{\bibinfo{person}{Andr{\'e}~P. Freire}, \bibinfo{person}{Raphael~W. {de Bettio}}, \bibinfo{person}{Elaine~G. Frade}, \bibinfo{person}{Fernanda~B. Ferrari}, \bibinfo{person}{Jos{\'e} Monserrat~Neto}, {and} \bibinfo{person}{Helena Libardi}.} \bibinfo{year}{2013}\natexlab{}.
  \newblock \showarticletitle{Accessibility of {{Web}} and {{Multimedia Content}}: {{Techniques}} and {{Examples}} from the {{Educational Context}}}. In \bibinfo{booktitle}{\emph{Proceedings of the 19th {{Brazilian Symposium}} on {{Multimedia}} and the {{Web}}}} \emph{(\bibinfo{series}{{{WebMedia}} '13})}. \bibinfo{publisher}{{ACM}}, \bibinfo{address}{{New York, NY, USA}}, \bibinfo{pages}{7--8}.
  \newblock
  \showISBNx{978-1-4503-2559-2}
  \urldef\tempurl%
  \url{https://doi.org/10.1145/2526188.2528538}
  \showDOI{\tempurl}
  
  
  \bibitem[Jia et~al\mbox{.}(2021)]%
          {jiaInfusingAccessibilityProgramming2021}
  \bibfield{author}{\bibinfo{person}{Lin Jia}, \bibinfo{person}{Yasmine~N. Elglaly}, \bibinfo{person}{Catherine~M. Baker}, {and} \bibinfo{person}{Kristen Shinohara}.} \bibinfo{year}{2021}\natexlab{}.
  \newblock \showarticletitle{Infusing {{Accessibility}} into {{Programming Courses}}}. In \bibinfo{booktitle}{\emph{Extended {{Abstracts}} of the 2021 {{CHI Conference}} on {{Human Factors}} in {{Computing Systems}}}}. \bibinfo{publisher}{{ACM}}, \bibinfo{address}{{Yokohama Japan}}, \bibinfo{pages}{1--6}.
  \newblock
  \showISBNx{978-1-4503-8095-9}
  \urldef\tempurl%
  \url{https://doi.org/10.1145/3411763.3451625}
  \showDOI{\tempurl}
  
  
  \bibitem[Kurniawan et~al\mbox{.}(2010)]%
          {kurniawanGeneralEducationCourse2010}
  \bibfield{author}{\bibinfo{person}{Sri~H. Kurniawan}, \bibinfo{person}{Sonia Arteaga}, {and} \bibinfo{person}{Roberto Manduchi}.} \bibinfo{year}{2010}\natexlab{}.
  \newblock \showarticletitle{A {{General Education Course}} on {{Universal Access}}, {{Disability}}, {{Technology}} and {{Society}}}. In \bibinfo{booktitle}{\emph{Proceedings of the 12th {{International ACM SIGACCESS Conference}} on {{Computers}} and {{Accessibility}}}} \emph{(\bibinfo{series}{{{ASSETS}} '10})}. \bibinfo{publisher}{{ACM}}, \bibinfo{address}{{New York, NY, USA}}, \bibinfo{pages}{11--18}.
  \newblock
  \showISBNx{978-1-60558-881-0}
  \urldef\tempurl%
  \url{https://doi.org/10.1145/1878803.1878808}
  \showDOI{\tempurl}
  
  
  \bibitem[Lewthwaite et~al\mbox{.}(2023)]%
          {lewthwaiteResearchingPedagogyDigital2023}
  \bibfield{author}{\bibinfo{person}{Sarah Lewthwaite}, \bibinfo{person}{Sarah Horton}, {and} \bibinfo{person}{Andy Coverdale}.} \bibinfo{year}{2023}\natexlab{}.
  \newblock \showarticletitle{Researching {{Pedagogy}} in {{Digital Accessibility Education}}}.
  \newblock \bibinfo{journal}{\emph{ACM SIGACCESS Accessibility and Computing}} \bibinfo{number}{134} (\bibinfo{date}{Jan.} \bibinfo{year}{2023}), \bibinfo{pages}{2:1}.
  \newblock
  \showISSN{1558-2337}
  \urldef\tempurl%
  \url{https://doi.org/10.1145/3582298.3582300}
  \showDOI{\tempurl}
  
  
  \bibitem[Lieby(2019)]%
          {liebyWorldwideProfessionalDeveloper2019}
  \bibfield{author}{\bibinfo{person}{Violet Lieby}.} \bibinfo{year}{2019}\natexlab{}.
  \newblock \bibinfo{title}{Worldwide {{Professional Developer Population}} of 24 {{Million Projected}} to {{Grow}} amid {{Shifting Geographical Concentrations}}}.
  \newblock \bibinfo{howpublished}{https://evansdata.com/press/viewRelease.php?pressID=278}.
  \newblock
  
  
  \bibitem[Ludi(2007)]%
          {ludiIntroducingAccessibilityRequirements2007}
  \bibfield{author}{\bibinfo{person}{Stephanie Ludi}.} \bibinfo{year}{2007}\natexlab{}.
  \newblock \showarticletitle{Introducing {{Accessibility Requirements}} through {{External Stakeholder Utilization}} in an {{Undergraduate Requirements Engineering Course}}}. In \bibinfo{booktitle}{\emph{29th {{International Conference}} on {{Software Engineering}} ({{ICSE}}'07)}}. \bibinfo{publisher}{{IEEE}}, \bibinfo{address}{{Minneapolis, MN, USA}}, \bibinfo{pages}{736--743}.
  \newblock
  \showISBNx{978-0-7695-2828-1}
  \showISSN{0270-5257}
  \urldef\tempurl%
  \url{https://doi.org/10.1109/ICSE.2007.46}
  \showDOI{\tempurl}
  
  
  \bibitem[Matausch et~al\mbox{.}(2006)]%
          {matauschAssistecUniversityCourse2006}
  \bibfield{author}{\bibinfo{person}{Kerstin Matausch}, \bibinfo{person}{Barbara Hengstberger}, {and} \bibinfo{person}{Klaus Miesenberger}.} \bibinfo{year}{2006}\natexlab{}.
  \newblock \showarticletitle{``{{Assistec}}'' \textendash{} {{A University Course}} on {{Assistive Technologies}}}.
  \newblock In \bibinfo{booktitle}{\emph{Computers {{Helping People}} with {{Special Needs}}}}, \bibfield{editor}{\bibinfo{person}{Klaus Miesenberger}, \bibinfo{person}{Joachim Klaus}, \bibinfo{person}{Wolfgang~L. Zagler}, \bibinfo{person}{Arthur~I. Karshmer}, \bibinfo{person}{David Hutchison}, \bibinfo{person}{Takeo Kanade}, \bibinfo{person}{Josef Kittler}, \bibinfo{person}{Jon~M. Kleinberg}, \bibinfo{person}{Friedemann Mattern}, \bibinfo{person}{John~C. Mitchell}, \bibinfo{person}{Moni Naor}, \bibinfo{person}{Oscar Nierstrasz}, \bibinfo{person}{C.~Pandu~Rangan}, \bibinfo{person}{Bernhard Steffen}, \bibinfo{person}{Madhu Sudan}, \bibinfo{person}{Demetri Terzopoulos}, \bibinfo{person}{Dough Tygar}, \bibinfo{person}{Moshe~Y. Vardi}, {and} \bibinfo{person}{Gerhard Weikum}} (Eds.). Vol.~\bibinfo{volume}{4061}. \bibinfo{publisher}{{Springer Berlin Heidelberg}}, \bibinfo{address}{{Berlin, Heidelberg}}, \bibinfo{pages}{361--368}.
  \newblock
  \showISBNx{978-3-540-36020-9 978-3-540-36021-6}
  \urldef\tempurl%
  \url{https://doi.org/10.1007/11788713_54}
  \showDOI{\tempurl}
  
  
  \bibitem[Patel et~al\mbox{.}(2020)]%
          {patelWhySoftwareNot2020}
  \bibfield{author}{\bibinfo{person}{Rohan Patel}, \bibinfo{person}{Pedro Breton}, \bibinfo{person}{Catherine~M. Baker}, \bibinfo{person}{Yasmine~N. {El-Glaly}}, {and} \bibinfo{person}{Kristen Shinohara}.} \bibinfo{year}{2020}\natexlab{}.
  \newblock \showarticletitle{Why {{Software}} Is {{Not Accessible}}: {{Technology Professionals}}' {{Perspectives}} and {{Challenges}}}. In \bibinfo{booktitle}{\emph{Extended {{Abstracts}} of the 2020 {{CHI Conference}} on {{Human Factors}} in {{Computing Systems}}}}. \bibinfo{publisher}{{ACM}}, \bibinfo{address}{{Honolulu HI USA}}, \bibinfo{pages}{1--9}.
  \newblock
  \showISBNx{978-1-4503-6819-3}
  \urldef\tempurl%
  \url{https://doi.org/10.1145/3334480.3383103}
  \showDOI{\tempurl}
  
  
  \bibitem[Rosmaita(2006)]%
          {rosmaitaAccessibilityFirstNew2006}
  \bibfield{author}{\bibinfo{person}{Brian~J. Rosmaita}.} \bibinfo{year}{2006}\natexlab{}.
  \newblock \showarticletitle{Accessibility {{First}}!: {{A New Approach}} to {{Web Design}}}. In \bibinfo{booktitle}{\emph{Proceedings of the 37th {{SIGCSE Technical Symposium}} on {{Computer Science Education}}}} \emph{(\bibinfo{series}{{{SIGCSE}} '06})}. \bibinfo{publisher}{{ACM}}, \bibinfo{address}{{New York, NY, USA}}, \bibinfo{pages}{270--274}.
  \newblock
  \showISBNx{978-1-59593-259-4}
  \urldef\tempurl%
  \url{https://doi.org/10.1145/1121341.1121426}
  \showDOI{\tempurl}
  
  
  \bibitem[Ross et~al\mbox{.}(2017)]%
          {rossAccessibilityFirstclassConcern2017}
  \bibfield{author}{\bibinfo{person}{Joel Ross}, \bibinfo{person}{Andrew~J. Ko}, {and} \bibinfo{person}{David~L. Stearns}.} \bibinfo{year}{2017}\natexlab{}.
  \newblock \showarticletitle{Accessibility as a First-Class Concern in Teaching {{GUIs}} and Software Engineering (Abstract Only)}. In \bibinfo{booktitle}{\emph{Proceedings of the 2017 {{ACM SIGCSE}} Technical Symposium on Computer Science Education}} \emph{(\bibinfo{series}{{{SIGCSE}} '17})}. \bibinfo{publisher}{{Association for Computing Machinery}}, \bibinfo{address}{{New York, NY, USA}}, \bibinfo{pages}{701}.
  \newblock
  \showISBNx{978-1-4503-4698-6}
  \urldef\tempurl%
  \url{https://doi.org/10.1145/3017680.3022393}
  \showDOI{\tempurl}
  
  
  \bibitem[Salda{\~n}a(2021)]%
          {saldanaCodingManualQualitative2021}
  \bibfield{author}{\bibinfo{person}{Johnny Salda{\~n}a}.} \bibinfo{year}{2021}\natexlab{}.
  \newblock \bibinfo{booktitle}{\emph{The Coding Manual for Qualitative Researchers} (\bibinfo{edition}{4th} ed.)}.
  \newblock \bibinfo{publisher}{{SAGE Publishing Ltd}}, \bibinfo{address}{{London [England] ; Thousand Oaks}}.
  \newblock
  \showISBNx{978-1-5297-3174-3 978-1-5297-3175-0}
  
  
  \bibitem[Shinohara et~al\mbox{.}(2018)]%
          {shinoharaWhoTeachesAccessibility2018}
  \bibfield{author}{\bibinfo{person}{Kristen Shinohara}, \bibinfo{person}{Saba Kawas}, \bibinfo{person}{Amy Ko}, {and} \bibinfo{person}{Richard~E. Ladner}.} \bibinfo{year}{2018}\natexlab{}.
  \newblock \showarticletitle{Who {{Teaches Accessibility}}? {{A Survey}} of {{U}}.{{S}}. {{Computing Faculty}}}. In \bibinfo{booktitle}{\emph{Proceedings of the 49th {{ACM Technical Symposium}} on {{Computer Science Education}}}} \emph{(\bibinfo{series}{{{SIGCSE}}'18})}. \bibinfo{publisher}{{Association for Computing Machinery}}, \bibinfo{address}{{Baltimore, Maryland, USA}}, \bibinfo{pages}{197--202}.
  \newblock
  \showISBNx{978-1-4503-5103-4}
  \urldef\tempurl%
  \url{https://doi.org/10.1145/3159450.3159484}
  \showDOI{\tempurl}
  
  
  \bibitem[Soares~Guedes and Landoni(2020)]%
          {soaresguedesHowAreWe2020}
  \bibfield{author}{\bibinfo{person}{Leandro Soares~Guedes} {and} \bibinfo{person}{Monica Landoni}.} \bibinfo{year}{2020}\natexlab{}.
  \newblock \showarticletitle{How {{Are We Teaching}} and {{Dealing}} with {{Accessibility}}? {{A Survey From Switzerland}}}. In \bibinfo{booktitle}{\emph{9th {{International Conference}} on {{Software Development}} and {{Technologies}} for {{Enhancing Accessibility}} and {{Fighting Info-exclusion}}}}. \bibinfo{publisher}{{ACM}}, \bibinfo{address}{{Online Portugal}}, \bibinfo{pages}{141--146}.
  \newblock
  \showISBNx{978-1-4503-8937-2}
  \urldef\tempurl%
  \url{https://doi.org/10.1145/3439231.3440610}
  \showDOI{\tempurl}
  
  
  \bibitem[Tseng et~al\mbox{.}(2022)]%
          {tsengExplorationIntegratingAccessibility2022}
  \bibfield{author}{\bibinfo{person}{Chia-En Tseng}, \bibinfo{person}{Seoung~Ho Jung}, \bibinfo{person}{Yasmine~N. Elglaly}, \bibinfo{person}{Yudong Liu}, {and} \bibinfo{person}{Stephanie Ludi}.} \bibinfo{year}{2022}\natexlab{}.
  \newblock \showarticletitle{Exploration on {{Integrating Accessibility}} into an {{AI Course}}}. In \bibinfo{booktitle}{\emph{Proceedings of the 53rd {{ACM Technical Symposium}} on {{Computer Science Education}}}}. \bibinfo{publisher}{{ACM}}, \bibinfo{address}{{Providence RI USA}}, \bibinfo{pages}{864--870}.
  \newblock
  \showISBNx{978-1-4503-9070-5}
  \urldef\tempurl%
  \url{https://doi.org/10.1145/3478431.3499399}
  \showDOI{\tempurl}
  \balance
  
  \bibitem[Wald(2008)]%
          {waldDesign10Credit2008}
  \bibfield{author}{\bibinfo{person}{Mike Wald}.} \bibinfo{year}{2008}\natexlab{}.
  \newblock \showarticletitle{Design of a 10 Credit Masters Level Assistive Technologies and Universal Design Module}. In \bibinfo{booktitle}{\emph{Proceedings of the 11th International Conference on {{Computers Helping People}} with {{Special Needs}}}} \emph{(\bibinfo{series}{{{ICCHP}} '08})}. {Springer}, \bibinfo{publisher}{{Springer-Verlag}}, \bibinfo{address}{{Berlin, Heidelberg}}, \bibinfo{pages}{190--193}.
  \newblock
  \showISBNx{978-3-540-70539-0}
  \urldef\tempurl%
  \url{https://doi.org/10.1007/978-3-540-70540-6_27}
  \showDOI{\tempurl}
  
  
  \bibitem[Walther and Ladner(2021)]%
          {waltherBroadeningParticipationTeaching2021}
  \bibfield{author}{\bibinfo{person}{Kendra Walther} {and} \bibinfo{person}{Richard~E. Ladner}.} \bibinfo{year}{2021}\natexlab{}.
  \newblock \showarticletitle{Broadening Participation by Teaching Accessibility}.
  \newblock \bibinfo{journal}{\emph{Commun. ACM}} \bibinfo{volume}{64}, \bibinfo{number}{10} (\bibinfo{date}{Oct.} \bibinfo{year}{2021}), \bibinfo{pages}{19--21}.
  \newblock
  \showISSN{0001-0782, 1557-7317}
  \urldef\tempurl%
  \url{https://doi.org/10.1145/3481356}
  \showDOI{\tempurl}
  
  
  \bibitem[Wang(2012)]%
          {wangHolisticPragmaticApproach2012}
  \bibfield{author}{\bibinfo{person}{Ye~Diana Wang}.} \bibinfo{year}{2012}\natexlab{}.
  \newblock \showarticletitle{A {{Holistic}} and {{Pragmatic Approach}} to {{Teaching Web Accessibility}} in an {{Undergraduate Web Design Course}}}. In \bibinfo{booktitle}{\emph{Proceedings of the 13th {{Annual Conference}} on {{Information Technology Education}}}} \emph{(\bibinfo{series}{{{SIGITE}} '12})}. \bibinfo{publisher}{{ACM}}, \bibinfo{address}{{New York, NY, USA}}, \bibinfo{pages}{55--60}.
  \newblock
  \showISBNx{978-1-4503-1464-0}
  \urldef\tempurl%
  \url{https://doi.org/10.1145/2380552.2380568}
  \showDOI{\tempurl}
  
  
  \bibitem[{World Health Organization}(2011)]%
          {worldhealthorganizationWorldReportDisability2011}
  \bibfield{author}{\bibinfo{person}{{World Health Organization}}.} \bibinfo{year}{2011}\natexlab{}.
  \newblock \bibinfo{title}{World {{Report}} on {{Disability Summary}}}.
  \newblock
  \newblock
  
  
  \bibitem[Yan and Ramachandran(2019)]%
          {yanCurrentStatusAccessibility2019}
  \bibfield{author}{\bibinfo{person}{Shunguo Yan} {and} \bibinfo{person}{P.~G. Ramachandran}.} \bibinfo{year}{2019}\natexlab{}.
  \newblock \showarticletitle{The {{Current Status}} of {{Accessibility}} in {{Mobile Apps}}}.
  \newblock \bibinfo{journal}{\emph{ACM Transactions on Accessible Computing}} \bibinfo{volume}{12}, \bibinfo{number}{1} (\bibinfo{date}{Feb.} \bibinfo{year}{2019}), \bibinfo{pages}{1--31}.
  \newblock
  \showISSN{1936-7228, 1936-7236}
  \urldef\tempurl%
  \url{https://doi.org/10.1145/3300176}
  \showDOI{\tempurl}

  \bibitem[Web(2023)]%
        {WebAIMWebAIMMillion2023}
 \bibinfo{year}{2023}\natexlab{}.
\newblock \bibinfo{title}{{{WebAIM}}: {{The WebAIM Million}} - {{The}} 2023 Report on the Accessibility of the Top 1,000,000 Home Pages}.
\newblock \bibinfo{howpublished}{https://webaim.org/projects/million/}.
\newblock

  \bibitem[Parthasarathy(2024)]%
          {parthasarathyExploringNeedAccessibility2024}
  \bibfield{author}{\bibinfo{person}{Parthasarathy PD} {and} \bibinfo{person}{Swaroop Joshi}.} \bibinfo{year}{2024}\natexlab{}.
  \newblock \showarticletitle{Exploring the {{Need}} of {{Accessibility Education}} in the {{Software Industry}}: {{Insights}} from a {{Survey}} of {{Software Professionals}} in {{India}}}.
  \newblock \bibinfo{journal}{\emph{46th {{International Conference}} on {{Software Engineering}} -- {{Software Engineering Education}} and {{Training Track}}}}.
  \newblock
  \urldef\tempurl%
  \url{http://arxiv.org/abs/2401.00451}
  \showDOI{\tempurl}

  \bibitem{10.1145/3664191}Kumar, A., Raj, R., Aly, S., Anderson, M., Becker, B., Blumenthal, R., Eaton, E., Epstein, S., Goldweber, M., Jalote, P., Lea, D., Oudshoorn, M., Pias, M., Reiser, S., Servin, C., Simha, R., Winters, T. \& Xiang, Q. Computer Science Curricula 2023. (Association for Computing Machinery,2024)


    \bibitem{edwardrajPerceptionsIntellectualDisability2010}Edwardraj, S., Mumtaj, K., Prasad, J., Kuruvilla, A. \& Jacob, K. Perceptions about Intellectual Disability: A Qualitative Study from Vellore, South India. {\em Journal Of Intellectual Disability Research}. \textbf{54}, 736-748 (2010)

    \bibitem[WorldBankDisabilityReport(2009)]%
        {WorldBankDisabilityReport2009}
    \bibinfo{year}{2009}\natexlab{}.
    \newblock \bibinfo{title}{{People with Disabilities in India: From commitments to outcomes}}.
    \newblock \bibinfo{howpublished}{World Bank, Human Development Unit, South Asia Region }.
    \newblock

    \bibitem{Das2013InclusiveEI}Das, A., Kuyini, A. \& Desai, I. Inclusive Education in India: Are the Teachers Prepared?.. {\em International Journal Of Special Education}. \textbf{28} pp. 27-36 (2013), https://api.semanticscholar.org/CorpusID:153642934

    \bibitem{TanejaJohansson2021EducationOC}Taneja-Johansson, S., Singal, N. \& Samson, M. Education of Children with Disabilities in Rural Indian Government Schools: A Long Road to Inclusion. {\em International Journal Of Disability, Development And Education}. \textbf{70} pp. 735 - 750 (2021), https://api.semanticscholar.org/CorpusID:236340430

    \bibitem{10.1371/journal.pone.0290016}Aruldas, K., Banks, L., Nagarajan, G., Roshan, R., Johnson, J., Musendo, D., Arpudharangam, I., Walson, J., Shakespeare, T. \& Ajjampur, S. “If He Has Education, There Will Not Be Any Problem”: Factors Affecting Access to Education for Children with Disabilities in Tamil Nadu, India. {\em PLOS ONE}. \textbf{18}, 1-19 (2023,8)

    \bibitem{ministryofsocialjusticeandempowermentgovernmentofindiaRightsPersonsDisabilities2016}Ministry of Social Justice and Empowerment, Government of India The Rights of Persons with Disabilities (RPD) Act, 2016.  (2016,12)

   \bibitem[Coverdale et~al\mbox{.}(2022)]%
          {coverdaleTeachingAccessibilityShared2022}
  \bibfield{author}{\bibinfo{person}{Coverdale, A.}, \bibinfo{person}{Lewthwaite, S.}, {and} \bibinfo{person}{Horton, S.}} \bibinfo{year}{2022}\natexlab{}.
  \newblock \showarticletitle{Teaching Accessibility as a Shared Endeavour: Building Capacity across Academic and Workplace Contexts.}. In \bibinfo{booktitle}{\emph{Proceedings of the 19th International Web For All Conference}} \bibinfo{publisher}{{Association for Computing Machinery}}, 
  \bibinfo{pages}{1--5}.
  \newblock
  \urldef\tempurl%
  \url{https://dl.acm.org/doi/10.1145/3493612.3520451}
  \showDOI{\tempurl}

   \bibitem[Elglaly et~al\mbox{.}(2024)]%
          {elglalyHCINeedAccessibility2024a}
  \bibfield{author}{\bibinfo{person}{Elglaly, Y.}, \bibinfo{person}{Baker, C.}, \bibinfo{person}{ Ross, A.} {and} \bibinfo{person}{Shinohara, K.}} \bibinfo{year}{2024}\natexlab{}.
  \newblock \showarticletitle{Beyond HCI: The Need for Accessibility Across the CS Curriculum.}  \bibinfo{booktitle}{\emph{Proceedings Of The 55th ACM Technical Symposium On Computer Science Education V. 1}} \bibinfo{publisher}{{Association for Computing Machinery}}, 
  \bibinfo{pages}{ pp. 324-330 (2024,3)}.
  \newblock
  \urldef\tempurl%
  \url{https://dl.acm.org/doi/10.1145/3626252.3630788}
  \showDOI{\tempurl}

   \bibitem[Oleson et~al\mbox{.}(2024)]%
          {TeachingAccessibleComputing2024}
  \bibfield{author}{\bibinfo{person}{Alannah Oleson}, \bibinfo{person}{Amy J. Ko}, {and} \bibinfo{person}{Richard Ladner}} \bibinfo{year}{2024}\natexlab{}.
  \newblock \showarticletitle{Teaching Accessible Computing}  \bibinfo{booktitle} 
  \bibinfo{publisher}{{Bookish Press}}, 
  \newblock
  \urldef\tempurl%
  \url{https://bookish.press/tac}
  \showDOI{\tempurl}

  \bibitem{10.1145/2968453} Ko, Amy J. \& Ladner, Richard E. AccessComputing Promotes Teaching Accessibility. {\em ACM Inroads}. \textbf{7}, 65-68 (2016,11), \url{https://doi.org/10.1145/2968453}

    
    
   \bibitem[Conn, P. et~al\mbox{.}(2024)]%
          {connUnderstandingMotivationsFinalyear2020}
  \bibfield{author}{\bibinfo{person}{Onn, P}, \bibinfo{person}{Gotfrid, T.}, \bibinfo{person}{Zhao, Q.}, \bibinfo{person}{Celestine, R.}, \bibinfo{person}{Mande, V.}, \bibinfo{person}{Shinohara, K.}, \bibinfo{person}{Ludi, S.} {and} \bibinfo{person}{Huenerfauth, M.}} \bibinfo{year}{2020}\natexlab{}.
  \newblock \showarticletitle{Understanding the Motivations of Final-year Computing Undergraduates for Considering Accessibility}  \bibinfo{booktitle} 
  \bibinfo{publisher}{{ACM Transactions On Computing Education}}, 
  \newblock
  \urldef\tempurl%
  \url{https://dl.acm.org/doi/10.1145/3381911}
  \showDOI{\tempurl}
    


  \end{thebibliography}
